\documentclass[preprint]{aastex62}
\usepackage{amsmath,amssymb}
\usepackage{longtable}
\usepackage{hyperref}
\usepackage{threeparttablex} 
\usepackage{textcomp}     
\usepackage[figuresright]{rotating}

\newcommand{\psj}{{Planetary Science Journal}}

\newcommand{\an}{{\textquotesingle Ayl\'{o}\textquotesingle chaxnim }}
\newcommand{\anns}{{\textquotesingle Ayl\'{o}\textquotesingle chaxnim}}

\newcommand{\nati}{{(163693) Atira }}
\newcommand{\natins}{{(163693) Atira}}

\received{--}
\revised{--}
\accepted{--}
\submitjournal{AJ}

\shorttitle{Atira's dark surface and origins}
\shortauthors{Bolin et al.}

\begin{document}

\title{The dark surface of (163693) Atira and implications for inner-Earth asteroid origins}

\author[0000-0002-4950-6323]{Bryce T. Bolin}
\affiliation{Eureka Scientific, Oakland, CA 94602, USA}
\correspondingauthor{Bryce Bolin}
\email{bbolin@eurekasci.com}

\author[0000-0002-7053-5495]{Carl Ingebretsen}
\affil{William H. Miller III Department of Physics and Astronomy, Johns Hopkins University, Baltimore, MD 21218, USA}

\author[0000-0001-7830-028X]{Robert Jedicke}
\affiliation{Institute for Astronomy, University of Hawai`i at M\={a}noa, Honolulu, HI 96822, USA}

\author[0000-0002-1804-7814]{William F. Bottke}
\affil{Solar System Science and Exploration Division, Southwest Research Institute, Boulder, CO 80302, USA}

\author[0000-0001-7335-1715]{Colin Orion Chandler}
\affil{Dept. of Astronomy \& the DiRAC Institute, University of Washington, Seattle, WA 98195, USA}

\author[0000-0001-6730-7857]{Rogerio Deienno}
\affil{Solar System Science and Exploration Division, Southwest Research Institute, Boulder, CO 80302, USA}

\author[0000-0001-8476-7687]{Alessandro Morbidelli}
\affil{Universit\'{e} C\^{o}te d'Azur, Observatoire de la C\^{o}te d'Azur, CNRS, Laboratoire Lagrange, 06300 Nice, France}
\affil{Coll\`{e}ge de France, 11 Pl. Berthelot, 75005 Paris, France}

\author[0000-0002-4547-4301]{David Nesvorn\'{y}}
\affil{Solar System Science and Exploration Division, Southwest Research Institute, Boulder, CO 80302, USA}

\author[0000-0002-2250-8687]{Leo Y. Alcorn}
\affil{W. M. Keck Observatory, Waimea, HI 96743, USA}

\author[0000-0003-0408-9850]{Percy Gomez}
\affil{W. M. Keck Observatory, Waimea, HI 96743, USA}

\author[0000-0001-7891-8143]{Meredith A. MacGregor}
\affil{William H. Miller III Department of Physics and Astronomy, Johns Hopkins University, Baltimore, MD 21218, USA}

\begin{abstract}
We compare the physical properties of inner-Earth asteroid (163693) Atira with the NEOMOD3 population model. Its radar-derived diameter of 4.9 km and absolute magnitude H$_V$$\simeq$16.4 imply a geometric albedo of p$_V$$\sim$0.02. Photometry with the Astrophysical Research Consortium 3.5-m telescope and ARCTIC yields g-r=0.49$\pm$0.03, r-i=0.20$\pm$0.02, i-z=0.02$\pm$0.03, a$^\ast$=$-0.05\pm0.03$, and a g-to-i spectral slope of 3.4$\pm$0.9$\%$/100 nm. The photometric reflectance favors a provisional X-type classification; combined with the low albedo, it supports a Tholen P-type interpretation. NEOMOD3 assigns an approximately 66$\%$ probability of delivery through inner Main Belt channels and 26$\%$ to central and outer Main Belt channels. Atira's NEOMOD3 albedo-category probabilities are $0.07\pm0.02$, $0.64\pm0.04$, and $0.29\pm0.04$ for p$_V$$<$0.1, 0.1$\leq$p$_V$$<$0.3, and p$_V$$\geq$0.3, respectively. Across 38 inner-Earth objects, including Atira and \anns, inner Main Belt delivery generally dominates, and intermediate albedo is favored for 33 objects. Atira therefore represents an uncommon low-albedo outcome, compatible with delivery of dark material from the inner Main Belt or more distant primitive populations. Further characterization of inner-Earth asteroids will test whether current population models reproduce their compositional diversity.
\end{abstract}

\section{Introduction}
Near-Earth objects (NEOs) are asteroids or comets with perihelion distances q$<$1.3 au \citep[][]{Binzel2015}, and can cross or share orbits with the terrestrial planets \citep[e.g.,][]{Granvik2013,Jedicke2025Lunar}. Atira asteroids, named after the prototype \natins, are inner-Earth objects with aphelia Q$<$0.983 au, while \textquotesingle Ayl\'{o}\textquotesingle chaxnim asteroids (Aylos) form a subset confined within Venus's orbit, with Q$<$0.718 au \citep[][]{Bolin2022IVO}. Their small maximum solar elongations, $<90^\circ$ for Atiras and $<46^\circ$ for Aylos, restrict ground-based searches to near-Sun twilight observations \citep[][]{Bolin2025TS}. Such surveys operate at low elevations, where high airmass, atmospheric extinction, and bright twilight backgrounds reduce sensitivity, making inner-Earth and inner-Venus asteroids difficult to detect and characterize \citep[][]{Schwamb2023ApJS}.

Although some NEOs may originate from tidal disruption events, lunar impacts, other inner solar system reservoirs, solar system comets, or interstellar comets \citep[][]{Bolin2020CD3,Bolin2023Com,Granvik2024,Jedicke2025Lunar,Alessi2026Lunar,Mommert2015NEOs,Bolin20253I,Rahatgaonkar20253IVLTgas}, population models indicate that the vast majority are supplied by resonances in the Main Belt \citep[][]{Nesvorny2023NEOMOD}. Models such as that of \citet[][]{Granvik2018} and NEOMOD \citep[][]{Nesvorny2023NEOMOD} describe the intrinsic NEO population and associate an object's semimajor axis, a, eccentricity, e, inclination, i, and absolute magnitude, H, with its probability of delivery through different Main Belt escape routes. These models identify the $\nu_6$ secular resonance and the 3:1 mean-motion resonance in the inner Main Belt as the most productive sources of NEOs \citep[e.g.,][]{Bottke2002a,Granvik2018}.

Although primitive, low-albedo asteroids such as C-types occur in the inner Main Belt, taxonomic surveys show that moderate-albedo, silicate-rich S- and Q-type asteroids are more common \citep[][]{DeMeo2014a}. Debiased compositional surveys of NEOs likewise indicate that objects delivered through the $\nu_6$ and 3:1 resonances are predominantly silicate-rich S- and Q-types \citep[][]{Binzel2019}. Studies of young asteroid families, such as Datura, constrain the size distribution, shapes, and spin states of fragments produced by Main Belt collisions \citep[][]{Vokrouhlicky2017Datura}. Some NEOs may originate from recent or intermediate-age asteroid-family-forming collisions, as spectroscopic and dynamical studies have suggested for the parent bodies of some L-chondrite meteorites \citep[][]{Marsset2024}.

While abundant ordinary-chondrite meteorites have been linked to silicate-rich asteroid families in the inner Main Belt, recent collisional and dynamical models have also associated carbonaceous chondrites and primitive NEOs with several inner Main Belt families \citep[][]{Broz2024Carbonaceous,Jenniskens2025}. The C-type NEOs Ryugu and Bennu, for example, likely originated from families in the inner Main Belt \citep[][]{Bottke2015YORPBennu,Delbo2023Bennu,Walsh2024Bennu}. Thus, although NEOs delivered from the inner Main Belt are expected to have predominantly silicate-rich compositions, the delivery of primitive asteroids represents a less common but plausible outcome.

\citet{Deleon2024} combined archival optical lightcurves with S-band Arecibo delay-Doppler radar imaging to derive a detailed shape model and mutual orbit for the known binary asteroid \nati (hereafter Atira). They found a primary volume-equivalent diameter of $4.92\pm0.95$ km, a secondary diameter of $0.8\pm0.3$ km, a mutual-orbit semimajor axis of $7.8\pm0.5$ km, and an orbital period of $15.577\pm0.003$ hr. These results imply a primary bulk density of $1.43\pm0.87$ g cm$^{-3}$. Their adopted absolute magnitude of H = 16.41 and radar-derived diameter imply a geometric albedo of p$_V$$\sim$0.02, substantially lower than the average NEO albedo of approximately 0.15 measured observationally and inferred from population models \citep[][]{Mainzer2011NEOprelim,Morbidelli2020albedo}.

In this work, we investigate Atira's taxonomic properties using complementary observational and statistical approaches. We present visible photometry of Atira obtained with the Astrophysical Research Consortium 3.5 m Telescope at Apache Point Observatory to constrain its surface composition and taxonomic classification. We compare its colors with the geometric albedo inferred from its published absolute magnitude and radar-derived size. We also use the NEO population model to calculate source-region and albedo-category probabilities for Atira and the known Atira population, placing its unusually dark surface in the broader context of the dynamical and physical properties of inner-Earth asteroids.

\section{Observations}
\label{s.obs}

Following the approach of \citet[][]{Bolin20253I} and \citet[][]{Ingebretsen2025RW1}, we observed Atira with the Astrophysical Research Consortium 3.5-m Telescope (ARC 3.5-m) at Apache Point Observatory on 2024 May 6 UT using the Astrophysical Research Consortium Telescope Imaging Camera (ARCTIC; \citealt{Huehnerhoff2016}). ARCTIC has a $7.85\arcmin \times 7.85\arcmin$ field of view and was operated using $2\times2$ pixel binning, quad-amplifier readout, and the medium readout rate, providing a pixel scale of 0.228 arcsec pixel$^{-1}$. Images were obtained using the Sloan Digital Sky Survey (SDSS) g, r, i, and z filters, which have effective wavelengths of 474 nm, 626 nm, 768 nm, and 963 nm \citep[][]{Fukugita1996,Huehnerhoff2016}.

We obtained 5 $\times$ 60 s exposures in each of the g, i, and z bands and 7 $\times$ 60 s exposures in r, giving total integration times of 300 s in each of g, i, and z and 420 s in r. We changed filters between individual exposures, interleaving the multiband measurements throughout the observing sequence \citep[e.g.,][]{Bolin20202I}. This cadence reduces the time separation between measurements in different bands but does not eliminate possible rotational biases in the colors. The telescope tracked at Atira's apparent rate of motion. The observations spanned an airmass range of 2.05--2.15 between 02:37:13 and 03:09:09 UT on 2024 May 6. This 31.9-min sequence covers approximately 16$\%$ of Atira's 3.398521-hr rotation period \citep{Deleon2024}; the residual color bias depends on the lightcurve amplitude and the temporal sampling in each filter. The seeing, measured from the full width at half-maximum of the r-band background stars, was 1.32 arcsec.

The images were bias-subtracted and flat-field corrected using calibration frames obtained in each filter using techniques of \citet[][]{Moreno2017J1}, \citet[][]{Bolin20181I}, and \citet[][]{Perley2025}. Photometric calibration was performed using field stars from the Pan-STARRS1 catalog \citep[][]{Chambers2016}, with the color transformations of \citet[][]{Tonry2012} used to convert the Pan-STARRS1 magnitudes to the SDSS photometric system \citep[e.g.,][]{Bolin2023,Bolin2025YR4}. Atira's photometry was measured in each image using a circular aperture with a radius of 2.05 arcsec. The sky background was measured within an annulus having inner and outer radii of 3.19 and 4.30 arcsec, respectively.

\section{Results}

\subsection{Photometry and spectral classification}
\label{sec:photo}
\begin{figure}
\centering
\includegraphics[scale=.405]{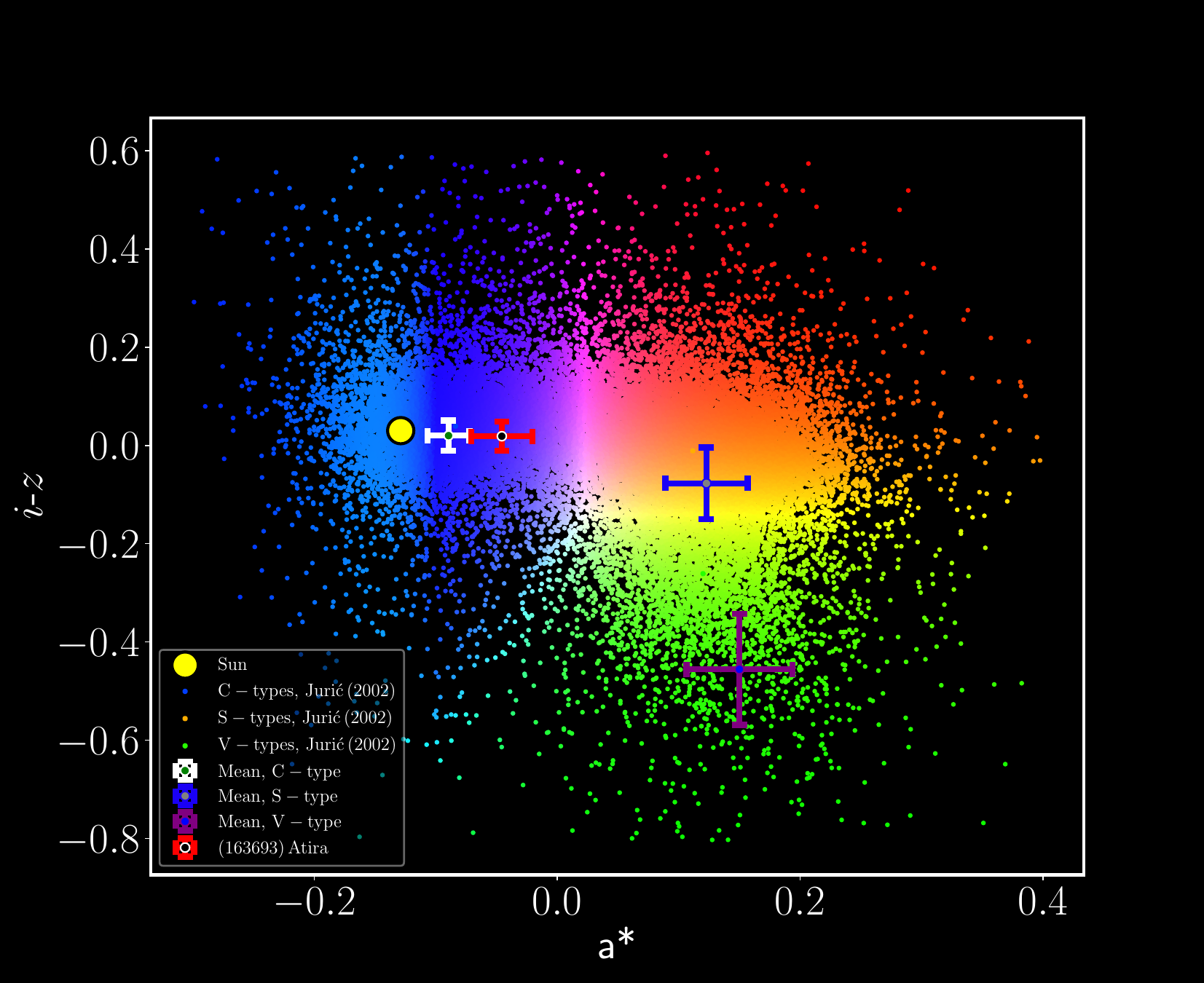}
\caption{Our calculated a$^{*}$ vs. i-z colors of Atira, plotted with a$^{*}$ vs. i-z colors of C- (blue), S- (red-orange) and V-type (green) asteroids observed in the Sloan Digital Sky Survey from \citet[][]{Ivezic2001} and \citet[][]{Juric2002}. The mean a$^{*}$ and $i$-$z$ C-, S-, and V-type asteroids are overplotted with one standard deviation error bars computed from the spectra of \citet[][]{DeMeo2009}. The location of the Sun at a$^{*}$ = -0.13 and i-z = 0.03 is plotted with a large black-bordered yellow dot.}
\label{fig:colors}
\end{figure}

From our photometry measurements of Atira, we measured magnitudes of g = 18.10 $\pm$ 0.02, r = 17.61 $\pm$ 0.01, i = 17.41 $\pm$ 0.02, z = 17.39 $\pm$ 0.03. The multi-band colors of Atira are g-r=0.49$\pm$0.03, r-i=0.20$\pm$0.02, g-i =0.69$\pm$0.03, and i-z=0.02$\pm$0.03. Using the definition for the principal color, a$^{*}$, from  \citet[][]{Ivezic2001}, a$^{*}$ = (0.89 (g-r)) + (0.45 (r-i)) - 0.57, and our g-r and r-i colors, we calculate an a$^{*}$ = -0.05$\pm$0.03. These principal and i-z colors place Atira closer to the C-type population than to the S- or V-type populations in a$^{*}$ vs. i-z space plotted in Fig.~\ref{fig:colors}. C-type asteroids have mean colors of a$^{*}$ = -0.09$\pm$0.02 and i-z = 0.02$\pm$0.03. Atira therefore has a slightly redder principal color, although consistent within the uncertainties, and the same mean i-z color. By comparison, S-types have mean colors of a$^{*}$ = 0.12$\pm$0.03 and i-z = -0.08$\pm$0.07, while V-types have a$^{*}$ = 0.15$\pm$0.11 and i-z = -0.46$\pm$0.04. Atira's colors are also consistent with the X complex, with representative values of a$^{*}$$\approx$-0.02 and i-z = 0.035$\pm$0.052 derived from the MANOS sample \citep[][]{Moskovitz2026}, illustrating the overlap between C- and X-complex asteroids in this color space. Compared with the Sun, which has a$^{*}$ = -0.13 and i-z = 0.03, Atira is redder in a$^{*}$ but has a consistent i-z color.

\begin{figure}
\centering
\includegraphics[scale=.405]{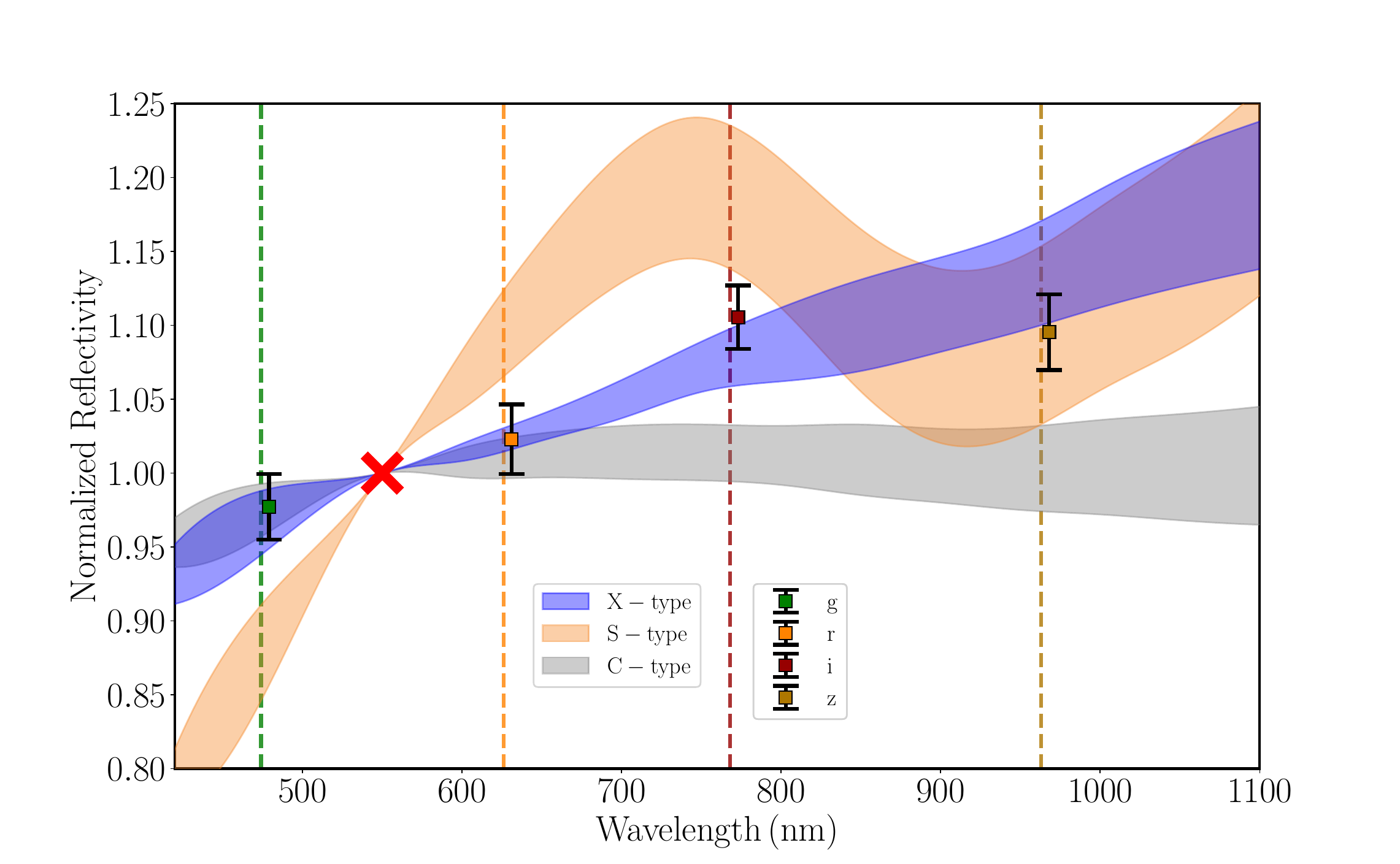}
\caption{Reflectance spectrophotometry of Atira derived from g, r, i, and z observations obtained with ARC 3.5-m/ARCTIC on 2024 May 6 UTC. Vertical dashed lines indicate the effective wavelengths of the filters, with the data points offset slightly in wavelength for clarity. Error bars represent 1$\sigma$ uncertainties. The reflectance is normalized to unity at 550 nm, indicated by the red X. Shaded regions show the spectral ranges of X-type (blue), S-type (orange), and C-type (gray) asteroids from the Bus-DeMeo taxonomic catalog \citep[][]{DeMeo2009}. Among the templates shown, the X-type spectrum most closely matches Atira's measured reflectance.}
\label{fig:reflectance}
\end{figure}

We derive Atira's spectral reflectance by dividing its measured flux in each g, r, i, and z filter by the corresponding solar-analog flux and normalizing the resulting reflectances at 550 nm. The spectrophotometry shown in Fig.~\ref{fig:reflectance} exhibits a modest red slope from g through i of 3.4$\pm0.9\%$/100 nm, followed by a slope between i and z of -0.5$\pm1.7\%$/100 nm. Previous photometry of Atira yielded a featureless spectrum compatible with X$_p$, D$_p$, or C$_p$ in the Carvano classification \citep{Rondon2022Atiras}. We compare these measurements with 20 asteroid taxonomic templates including S-complex asteroid types (S, Sa, Sq, Sr, Sv, Q),  C-complex types (B, C, Cg, Cgh), X-complex types (X, Xe, Xn), and other assorted types (A, D, K, L, O, R, V) \citep[][]{Bus2002,DeMeo2009} using the reduced $\chi^2$ statistic. The closest match was the X-type with a reduced  $\chi^2$ of 1.42. By comparison, the C-, S-, and V-type templates yield reduced $\chi^2$ values of 10.52, 14.46, and 62.80, respectively. Atira's low visible geometric albedo of p$_V$ $\sim0.02$, inferred from its radar-derived diameter and absolute magnitude \citep[][]{Deleon2024}, is consistent with the low-albedo members of X-types \citep[][]{Fornasier2011,DeMeo2013aa}, called P-types in the Tholen-Barucci asteroid taxonomy \citep[][]{Tholen1989}, favoring a provisional X-/P-type classification for Atira.

\subsection{NEO model source and albedo probability calculations}

We used NEOMOD3 \citep[][]{Nesvorny2023NEOMOD,Nesvorny2024NEOMOD3} to estimate the dynamical source probabilities of Atira and the other asteroids in our sample. NEOMOD3 combines the debiased orbital and absolute-magnitude distribution of NEOs with an albedo distribution constrained by NEOWISE observations. For each asteroid, we supplied its semimajor axis a, eccentricity e, inclination i, and absolute magnitude H, and retrieved the source probabilities directly from the corresponding model bin. For Atira, we adopted a = 0.741 au, e = 0.3221, i = $25.62^\circ$, and H = 16.41 mag. The model includes 12 source channels: the $\nu_6$ secular resonance; the 3:1, 5:2, 7:3, 8:3, 9:4, 11:5, and 2:1 mean-motion resonances with Jupiter; weak resonances in the inner Main Belt; the Hungaria and Phocaea populations; and a cometary component. 

Table~\ref{tab:orbital_source_probabilities} lists the source probabilities evaluated at each asteroid's a, e, i, and H. Atira's leading channels are the 3:1 resonance, weak inner Main Belt resonances, the $\nu_6$ resonance, and the 11:5 resonance. We also include (3200) Phaethon, (155140) 2005 UD, and (225416) 1999 YC for comparison in the following section.

We estimated albedo-category probabilities using a synthetic NEO population drawn from NEOMOD3. For each asteroid, we defined a selection window extending $\pm0.1$ au in a, $\pm0.05$ in e, $\pm5^\circ$ in i, and $\pm0.5$ mag in H around its adopted values. The initial realization contained only 12 objects within Atira's window, so we increased the synthetic population by a factor of 100 to improve the sampling statistics. The objects selected within each window provide the model's expected albedo distribution conditional on orbital elements and absolute magnitude. We divided the selected objects into three categories according to their visible geometric albedo: low albedo, p$_V$ $<0.1$; intermediate albedo, $0.1\leq$ p$_V$ $<0.3$; and high albedo, p$_V$ $\geq0.3$. The probability of each category was calculated as the fraction of selected synthetic objects within that albedo interval. For Atira, the resulting probabilities were $0.07\pm0.02$, $0.64\pm0.04$, and $0.29\pm0.04$ for the low-, intermediate-, and high-albedo categories, respectively, with uncertainties at the $1\sigma$ level. Table~\ref{tab:albedo_categories} presents the albedo-category probabilities for Atira, the other inner-Earth asteroids, and the comparison objects (3200) Phaethon, (155140) 2005 UD, and (225416) 1999 YC.

\section{Discussion and Conclusions}

Atira's photometry favors a provisional X-type classification, while its radar-derived size and absolute magnitude imply a very low albedo \citep{Deleon2024}. Together, these constraints support a P-type interpretation and motivate considering both an inner Main Belt source and delivery from the primitive populations of the central and outer Main Belt. Our NEOMOD3 results assign a combined probability of approximately 66$\%$ to delivery through the 3:1 resonance, weak inner Main Belt resonances, and the $\nu_6$ resonance. This probability describes escape channels but does not uniquely identify the earlier location of Atira's parent body: the 3:1 resonance can also receive objects evolving inward from the central Main Belt \citep[][]{Delbo2019Ancient,Nesvorny2021DarkKT}.

Ancient inner Main Belt families could supply X-complex objects to near-Earth space. \citet{Delbo2019Ancient} identified the approximately 3-Gyr-old Athor family and potentially primordial Zita family, whose fragments can reach the $\nu_6$, 3:1, and weaker escape resonances through thermal drift. These families illustrate a possible delivery mechanism, although their moderate albedos, $0.1<$ p$_V$ $<0.3$, exceed Atira's and neither is established as its parent. However, the limited size-dependent orbital dispersal of the dark Erigone family does not support delivery of an Atira-sized fragment to near-Earth space \citep[][their Figure~2]{Marchi2025Donaldjohanson}. The same constraint applies to Clarissa \citep[][their Figure~5]{Bottke2015YORPBennu}. We therefore exclude these two families as sources within the published family-evolution models, while retaining the possibility of another primitive inner Main Belt source.

Delivery from the central or outer Main Belt is also plausible. In the optimized-drift model of \citet{Nesvorny2021DarkKT}, approximately 71$\%$ of dark near-Earth asteroids larger than 5 km originate beyond 2.5 au. Their simulations also attribute approximately 60$\%$ of dark impactors on the terrestrial planets to these regions, providing a pathway for primitive Cretaceous--Paleogene (K/Pg)-scale impactors. Atira's diameter of $4.92\pm0.95$ km lies near the lower size boundary of these separate large-NEA simulations, which followed D$>$5 km asteroids. This restriction does not apply to NEOMOD3 \citep{Nesvorny2024NEOMOD3}. The large-NEA population fractions therefore provide context for Atira's possible origin, rather than an individual source probability. Our NEOMOD3 analysis gives Atira a combined probability of approximately 26$\%$ for delivery through the central and outer Main Belt 8:3, 5:2, 7:3, 9:4, 11:5, and 2:1 resonances, dominated by the 11:5 contribution of approximately 19$\%$. The prevalence of low-albedo P-types in the outer Main Belt \citep[][]{DeMeo2014a,Vokrouhlicky2016Capture} is consistent with, but does not uniquely establish, an origin there for Atira.

Atira could also have a distant primordial origin. \citet[][]{Vokrouhlicky2016Capture} showed that the giant-planet instability could implant planetesimals from a primordial trans-Neptunian disk throughout the Main Belt, preferentially in its outer regions. These bodies or their collisional fragments could remain there for billions of years before escaping through resonances into near-Earth space. Atira could represent material that experienced two distinct episodes of transport: primordial implantation into the Main Belt, followed much later by delivery into its present inner-Solar-System orbit. This possibility is also compatible with an inner Main Belt source since some of the implanted objects in the model described by \citet[][]{Vokrouhlicky2016Capture} could have been implanted there during the giant planet instability.

Atira, Phaethon, and the proposed Phaethon-associated object 2005 UD have inclinations of $25.6^\circ$, $22.31^\circ$, and $28.61^\circ$, respectively, within the favored interval of $5^\circ<$i$<30^\circ$ in the large-NEA delivery simulations of \citet{Nesvorny2021DarkKT}. Another object proposed to be associated with Phaethon, 1999 YC, lies above this interval at $38.26^\circ$. The inclination similarity is compatible with Atira and Phaethon belonging to the modeled large-NEA population. Their overlapping NEOMOD3 source distributions also support the possibility of a common escape channel, as discussed below, without establishing a shared parent body or asteroid family (Table~\ref{tab:orbital_source_probabilities}). All four have broadly primitive spectral properties \citep[][]{DeMeo2009}. Atira's modestly red spectrophotometry favors X-/P-types, Phaethon has a blue, largely featureless B-type spectrum \citep[][]{Kareta2018}, visible-to-near-infrared observations classify 2005 UD as Cb-type \citep[][]{Gherase2024}, and 1999 YC has approximately neutral optical colors consistent with C-types \citep[][]{KasugaJewitt2008}.

Phaethon and its proposed associated objects have low to moderate albedo estimates. For Phaethon, \citet{Kareta2018} obtained p$_V$ $=0.08\pm0.01$. However, recent modeling of NEOWISE data suggests that the albedo of Phaethon may be as large as $\sim$0.16 \cite[][]{Masiero2019}.  Asteroid 2005 UD has p$_V$ $=0.14\pm0.09$ from thermal-emission modeling \citep[][]{Masiero2019} and $0.10\pm0.02$ from polarimetry \citep[][]{Devogele2020}, and 1999 YC has p$_V$ $=0.09\pm0.03$  \citep[][]{Kasuga2022}. Although the albedo estimates depend on the observational and modeling methods, they place these objects among relatively dark to moderate-albedo NEAs. This broad similarity is compatible with several source populations, but does not establish a common dynamical origin.

Our NEOMOD3 source estimates favor inner Main Belt-associated delivery for Atira and the Phaethon comparison objects, with combined probabilities of approximately 66$\%$ for Atira and 76--84$\%$ for the comparison objects (Table~\ref{tab:orbital_source_probabilities}). Phaethon's distribution is dominated by weak inner Main Belt resonances, whereas the 3:1 resonance is the leading individual channel for Atira, 2005 UD, and 1999 YC. These overlapping distributions allow shared escape channels without establishing a common parent population.

We group the 5:2 and 8:3 resonances with Jupiter as central Main Belt-associated escape channels, and the 7:3, 9:4, 11:5, and 2:1 resonances as outer Main Belt-associated channels. We treat the Hungaria and Phocaea populations separately. Atira has the largest combined central and outer Main Belt contribution among these four objects, dominated by the 11:5 resonance, which also contributes appreciably to 2005 UD (Table~\ref{tab:orbital_source_probabilities}). These probabilities characterize delivery routes rather than unique formation locations, since central Main Belt material can also escape through the 3:1 resonance \citep{Nesvorny2021DarkKT}.

The NEOMOD3 source probabilities for our 38 inner-Earth objects generally favor inner Main Belt sources (Table~\ref{tab:orbital_source_probabilities}). The $\nu_6$ resonance has the highest individual probability for 27 objects, followed by the 11:5 resonance and weak inner Main Belt resonances for four objects each, and the 3:1 resonance for three objects. Averaging with equal weight for each object, the three inner Main Belt-associated channels contribute approximately 83.5$\%$, compared with 3.8$\%$ and 8.3$\%$ for central and outer Main Belt-associated channels, respectively.

The 11:5 resonance is the highest-probability individual source for 2008 EA$_{32}$, 2019 LF$_6$, 2020 OV$_1$, and 2008 UL$_{90}$, with probabilities ranging from 36$\%$ to 50$\%$, all exceeding Atira's. Only 2008 EA$_{32}$ has an outer Main Belt-associated probability exceeding the combined inner Main Belt contribution. These differences indicate that similar present-day confinement within Earth's orbit can be reached through several Main Belt delivery pathways.

The albedo-category probabilities in Table~\ref{tab:albedo_categories} favor intermediate albedo, 0.1$\leq$p$_V$$<$0.3, for 33 of the 38 inner-Earth objects. Low albedo, p$_V$$<$0.1, is favored for 1998 DK$_{36}$, 2017 XA$_1$, and 2024 WD$_{19}$, while high albedo, p$_V$$\geq$0.3, is favored for 2018 JB$_3$. Asteroid 2020 HA$_{10}$ has equal intermediate- and high-albedo probabilities at the reported precision. Averaging the individual distributions gives approximately 53.4$\%$ in the intermediate-albedo category, 26.0$\%$ in the high-albedo category, and 20.7$\%$ in the low-albedo category. These averages differ from the fractions of objects whose highest-probability category falls in each interval, because other categories remain plausible for each object.

The clearest low-albedo predictions are for 2017 XA$_1$ and 2024 WD$_{19}$, with probabilities of $0.68\pm0.01$ and $0.68\pm0.03$, respectively. The nominal low-albedo probability for 1998 DK$_{36}$ is $1.00^{+0.00}_{-0.43}$, although its large uncertainty limits the interpretation of this extreme value. In contrast, 2018 JB$_3$ favors the high-albedo category with a probability of $0.51\pm0.04$, while 2020 HA$_{10}$ has intermediate- and high-albedo probabilities of $0.44\pm0.04$ each. All quoted uncertainties are $1\sigma$. Atira itself follows the more common model preference for intermediate albedo, with a probability of $0.64\pm0.04$ and a weighted model albedo of $\tilde p_V=0.22$, despite its independently inferred low geometric albedo. This comparison emphasizes that the NEOMOD3 albedo probabilities are predictions conditioned on orbital elements and absolute magnitude, rather than direct measurements of individual surfaces.

Our ARC 3.5-m/ARCTIC g, r, i, and z photometry favors a provisional X-type classification for Atira. Its radar-derived size and absolute magnitude imply p$_V$ $\sim$0.02, supporting a P-type interpretation. NEOMOD3 assigns only 7$\pm$2$\%$ probability to p$_V$ $<$0.1 and favors inner Main Belt delivery (66$\%$), with central and outer Main Belt channels contributing 26$\%$. Across 38 inner-Earth objects, inner Main Belt delivery generally dominates, and intermediate albedo ranks highest for 33 objects; this count does not give the fraction of surfaces with intermediate albedos. Atira is an uncommon but allowed low-albedo outcome in NEOMOD3; this single object does not establish a discrepancy with the model. Its properties remain compatible with either a dark inner Main Belt source or more distant primitive populations, without identifying a unique parent. Further spectroscopy and albedo constraints will test whether dark inner-Earth asteroids occur at the frequency predicted by NEO population models.

\section*{Acknowledgments}
This research benefited from B. T. Bolin's and R. Jedicke's participation in the 2026 workshop ``Exploring Planetary Systems in the Era of Time-domain Astronomy'', hosted by the Institute for Astronomy at the University of Hawai'i and supported by award 2106927 from the National Science Foundation's Astronomy \& Astrophysics Research Grants program.

C.O.C. is supported by Schmidt Sciences. C.O.C. gratefully acknowledges support from the NASA CSSFP (grant No. 80NSSC26K0380). C.O.C. acknowledges support from the DiRAC Institute in the Department of Astronomy at the University of Washington. The DiRAC Institute is supported through generous gifts from the Charles and Lisa Simonyi Fund for Arts and Sciences, Janet and Lloyd Frink, and the Washington Research Foundation.

\facility{ARC} 

\bibliographystyle{aasjournal}

\begin{table}
\centering
\caption{Orbital properties and NEOMOD3 source probabilities for the 38 inner-Earth objects and three Phaethon-like comparison objects.}
\label{tab:orbital_source_probabilities}
\scriptsize
\setlength{\tabcolsep}{3pt}
\renewcommand{\arraystretch}{1.05}
\begin{tabular}[c]{lccccccccccccc}
\hline
Object & a & e & i & H & p$_{\nu_6}$ & p$_{3:1}$ & p$_{5:2}$ & p$_{8:3}$ & p$_{11:5}$ & p$_{\mathrm{inner}}$ & p$_{\mathrm{Hun}}$ & p$_{\mathrm{Pho}}$ & p$_{\mathrm{other}}$ \\
& (au) & & (deg) & & & & & & & & & & \\
\hline
\textbf{(163693) Atira} & 0.7410 & 0.3221 & 25.62 & 16.41 & 0.196 & 0.244 & 0.020 & 0.053 & 0.188 & 0.224 & 0.066 & 0.008 & 0.001 \\
(164294) 2004 XZ$_{130}$ & 0.6176 & 0.4545 & 2.950 & 20.45 & 0.636 & 0.192 & 0.013 & 0.017 & 0.044 & 0.064 & 0.034 & 0.001 & 0.000 \\
(413563) 2005 TG$_{45}$ & 0.6813 & 0.3723 & 23.33 & 17.63 & 0.342 & 0.182 & 0.013 & 0.005 & 0.173 & 0.203 & 0.076 & 0.007 & 0.000 \\
2008 EA$_{32}$ & 0.6159 & 0.3050 & 28.26 & 16.49 & 0.151 & 0.154 & 0.018 & 0.026 & 0.496 & 0.086 & 0.070 & 0.000 & 0.000 \\
2004 JG$_{6}$ & 0.6352 & 0.5311 & 18.94 & 18.48 & 0.416 & 0.149 & 0.072 & 0.058 & 0.102 & 0.164 & 0.035 & 0.005 & 0.000 \\
2008 UL$_{90}$ & 0.6950 & 0.3799 & 24.31 & 18.48 & 0.325 & 0.110 & 0.028 & 0.014 & 0.357 & 0.110 & 0.052 & 0.004 & 0.000 \\
(594913) \an & 0.5553 & 0.1772 & 15.87 & 16.20 & 0.255 & 0.286 & 0.000 & 0.055 & 0.000 & 0.292 & 0.101 & 0.000 & 0.011 \\
2006 WE$_{4}$ & 0.7847 & 0.1830 & 24.77 & 18.88 & 0.463 & 0.257 & 0.031 & 0.008 & 0.135 & 0.073 & 0.027 & 0.004 & 0.001 \\
2017 YH & 0.6346 & 0.4823 & 19.85 & 18.10 & 0.410 & 0.113 & 0.072 & 0.040 & 0.101 & 0.202 & 0.052 & 0.009 & 0.000 \\
1998 DK$_{36}$ & 0.6923 & 0.4160 & 2.020 & 25.01 & 0.805 & 0.126 & 0.008 & 0.002 & 0.000 & 0.051 & 0.007 & 0.001 & 0.000 \\
2010 XB$_{11}$ & 0.6181 & 0.5338 & 29.88 & 19.69 & 0.540 & 0.302 & 0.017 & 0.032 & 0.008 & 0.061 & 0.040 & 0.000 & 0.000 \\
2012 VE$_{46}$ & 0.7131 & 0.3612 & 6.670 & 20.20 & 0.599 & 0.183 & 0.011 & 0.028 & 0.022 & 0.121 & 0.031 & 0.004 & 0.000 \\
2013 JX$_{28}$ & 0.6008 & 0.5642 & 10.77 & 20.10 & 0.584 & 0.210 & 0.031 & 0.024 & 0.000 & 0.122 & 0.027 & 0.002 & 0.000 \\
2013 TQ$_{5}$ & 0.7737 & 0.1557 & 16.40 & 19.87 & 0.468 & 0.212 & 0.011 & 0.009 & 0.000 & 0.254 & 0.045 & 0.001 & 0.001 \\
2014 FO$_{47}$ & 0.7522 & 0.2711 & 19.20 & 20.21 & 0.519 & 0.266 & 0.014 & 0.008 & 0.002 & 0.161 & 0.028 & 0.001 & 0.001 \\
2015 DR$_{215}$ & 0.6666 & 0.4715 & 4.060 & 20.51 & 0.643 & 0.225 & 0.017 & 0.005 & 0.042 & 0.037 & 0.026 & 0.004 & 0.000 \\
2017 XA$_{1}$ & 0.8094 & 0.2016 & 17.18 & 21.30 & 0.531 & 0.203 & 0.013 & 0.009 & 0.000 & 0.224 & 0.018 & 0.002 & 0.000 \\
2018 JB$_{3}$ & 0.6832 & 0.2904 & 40.39 & 17.67 & 0.187 & 0.173 & 0.027 & 0.000 & 0.006 & 0.553 & 0.053 & 0.000 & 0.003 \\
2019 AQ$_{3}$ & 0.5886 & 0.3143 & 47.22 & 17.48 & 0.178 & 0.046 & 0.001 & 0.000 & 0.000 & 0.757 & 0.017 & 0.000 & 0.000 \\
2019 LF$_{6}$ & 0.5554 & 0.4293 & 29.50 & 17.28 & 0.211 & 0.154 & 0.003 & 0.006 & 0.406 & 0.170 & 0.050 & 0.000 & 0.000 \\
2020 HA$_{10}$ & 0.8197 & 0.1553 & 49.65 & 18.86 & 0.356 & 0.394 & 0.025 & 0.024 & 0.000 & 0.130 & 0.067 & 0.003 & 0.001 \\
2020 OV$_{1}$ & 0.6376 & 0.2541 & 32.58 & 18.72 & 0.334 & 0.113 & 0.026 & 0.008 & 0.365 & 0.085 & 0.069 & 0.000 & 0.000 \\
2021 BS$_{1}$ & 0.5984 & 0.3376 & 31.73 & 18.63 & 0.439 & 0.139 & 0.014 & 0.001 & 0.273 & 0.053 & 0.080 & 0.000 & 0.000 \\
2021 LJ$_{4}$ & 0.6764 & 0.3820 & 9.820 & 20.11 & 0.595 & 0.195 & 0.005 & 0.011 & 0.024 & 0.128 & 0.035 & 0.005 & 0.000 \\
2021 PB$_{2}$ & 0.7174 & 0.1500 & 24.83 & 18.80 & 0.507 & 0.248 & 0.038 & 0.010 & 0.098 & 0.056 & 0.038 & 0.003 & 0.000 \\
2021 PH$_{27}$ & 0.4618 & 0.7115 & 31.94 & 17.67 & 0.322 & 0.164 & 0.132 & 0.012 & 0.005 & 0.358 & 0.007 & 0.000 & 0.000 \\
2021 VR$_{3}$ & 0.5339 & 0.4139 & 18.07 & 18.01 & 0.488 & 0.244 & 0.019 & 0.017 & 0.012 & 0.158 & 0.058 & 0.001 & 0.002 \\
2022 BJ$_{8}$ & 0.7853 & 0.2487 & 15.83 & 19.41 & 0.421 & 0.216 & 0.011 & 0.014 & 0.000 & 0.307 & 0.032 & 0.000 & 0.001 \\
2023 EL & 0.7714 & 0.2462 & 13.88 & 19.17 & 0.405 & 0.218 & 0.011 & 0.014 & 0.000 & 0.317 & 0.033 & 0.000 & 0.001 \\
2023 EY$_{2}$ & 0.6036 & 0.3398 & 35.47 & 19.84 & 0.517 & 0.173 & 0.011 & 0.011 & 0.167 & 0.053 & 0.067 & 0.000 & 0.000 \\
2023 KQ$_{5}$ & 0.8722 & 0.07710 & 67.43 & 18.76 & 0.719 & 0.226 & 0.000 & 0.006 & 0.000 & 0.000 & 0.042 & 0.007 & 0.000 \\
2023 WK$_{3}$ & 0.6442 & 0.4998 & 24.47 & 20.46 & 0.636 & 0.186 & 0.002 & 0.042 & 0.038 & 0.049 & 0.044 & 0.002 & 0.000 \\
2024 UM$_{9}$ & 0.8609 & 0.06720 & 21.17 & 20.64 & 0.402 & 0.365 & 0.006 & 0.012 & 0.000 & 0.183 & 0.031 & 0.000 & 0.000 \\
2024 WD$_{19}$ & 0.6795 & 0.1579 & 3.790 & 21.32 & 0.619 & 0.207 & 0.012 & 0.006 & 0.000 & 0.130 & 0.026 & 0.000 & 0.000 \\
2025 GN$_{1}$ & 0.4620 & 0.7051 & 32.84 & 20.06 & 0.332 & 0.516 & 0.000 & 0.021 & 0.000 & 0.130 & 0.001 & 0.000 & 0.000 \\
2025 LP & 0.7437 & 0.1624 & 33.34 & 19.38 & 0.455 & 0.343 & 0.035 & 0.030 & 0.068 & 0.026 & 0.039 & 0.004 & 0.000 \\
2025 SC$_{79}$ & 0.4958 & 0.4499 & 16.77 & 18.83 & 0.462 & 0.279 & 0.005 & 0.001 & 0.000 & 0.213 & 0.039 & 0.000 & 0.000 \\
2026 BX$_{4}$ & 0.8750 & 0.1220 & 19.97 & 20.96 & 0.455 & 0.270 & 0.016 & 0.011 & 0.000 & 0.211 & 0.033 & 0.004 & 0.000 \\
\hline
\textbf{(3200) Phaethon} & 1.2710 & 0.8897 & 22.31 & 14.38 & 0.068 & 0.301 & 0.017 & 0.058 & 0.001 & 0.474 & 0.052 & 0.028 & 0.002 \\
\textbf{(155140) 2005 UD} & 1.2750 & 0.8724 & 28.61 & 17.47 & 0.258 & 0.284 & 0.019 & 0.020 & 0.102 & 0.272 & 0.033 & 0.010 & 0.000 \\
\textbf{(225416) 1999 YC} & 1.4220 & 0.8302 & 38.26 & 17.36 & 0.237 & 0.340 & 0.067 & 0.025 & 0.014 & 0.187 & 0.051 & 0.073 & 0.007 \\

\hline
\end{tabular}\par
\vspace{0.5em}
\begin{minipage}{0.98\linewidth}
\scriptsize
\textbf{Notes.}
a, e, i, and H are the semimajor axis, eccentricity, inclination, and absolute magnitude.
Source probabilities refer to the indicated resonances, weak inner Main Belt resonances (inner), Hungarias (Hun), and Phocaeas (Pho).
The other category sums the 7:3, 9:4, and 2:1 resonances and the Jupiter-family-comet source; each of these channels has a probability below 0.02 for every object in the table.
The full, uncombined probabilities are supplied in \texttt{full\_source\_probabilities.csv}.
Probabilities are retained to three decimal places for reproducibility and may not sum to unity because of rounding.
The three comparison objects appear below the horizontal rule.
\end{minipage}
\end{table}

\begin{table}
\centering
\caption{Albedo-category probabilities and weighted model albedos for the currently known Atira asteroids and Phaethon-like objects.}
\label{tab:albedo_categories}
\tiny
\setlength{\tabcolsep}{5pt}
\renewcommand{\arraystretch}{1.15}
\begin{tabular}{lcccc}
\hline
Object & Low albedo$^{1}$ & Medium albedo$^{2}$ & High albedo$^{3}$ &  $\tilde p_V$$^{4}$ \\
\hline
\textbf{(163693) Atira} & $0.07^{+0.02}_{-0.02}$ & $0.64^{+0.04}_{-0.04}$ & $0.29^{+0.04}_{-0.04}$ & $0.219^{+0.014}_{-0.014}$/$\sim$0.02$^*$ \\
(164294) 2004 XZ$_{130}$ & $0.25^{+0.02}_{-0.02}$ & $0.70^{+0.02}_{-0.02}$ & $0.05^{+0.01}_{-0.01}$ & $0.167^{+0.005}_{-0.005}$ \\
(413563) 2005 TG$_{45}$ & $0.15^{+0.02}_{-0.02}$ & $0.51^{+0.02}_{-0.02}$ & $0.34^{+0.02}_{-0.02}$ & $0.212^{+0.007}_{-0.007}$ \\
2008 EA$_{32}$ & $0.10^{+0.03}_{-0.03}$ & $0.57^{+0.05}_{-0.05}$ & $0.33^{+0.05}_{-0.05}$ & $0.218^{+0.018}_{-0.018}$ \\
2004 JG$_{6}$ & $0.13^{+0.01}_{-0.01}$ & $0.50^{+0.02}_{-0.02}$ & $0.37^{+0.02}_{-0.02}$ & $0.218^{+0.007}_{-0.007}$ \\
2008 UL$_{90}$ & $0.12^{+0.01}_{-0.01}$ & $0.50^{+0.02}_{-0.02}$ & $0.38^{+0.02}_{-0.02}$ & $0.220^{+0.007}_{-0.007}$ \\
(594913) \an & $0.14^{+0.07}_{-0.06}$ & $0.64^{+0.08}_{-0.09}$ & $0.21^{+0.08}_{-0.07}$ & $0.198^{+0.029}_{-0.028}$/$\sim$0.2$^{**}$ \\
2006 WE$_{4}$ & $0.11^{+0.01}_{-0.01}$ & $0.52^{+0.01}_{-0.01}$ & $0.37^{+0.01}_{-0.01}$ & $0.221^{+0.004}_{-0.004}$ \\
2017 YH & $0.13^{+0.02}_{-0.01}$ & $0.47^{+0.02}_{-0.02}$ & $0.41^{+0.02}_{-0.02}$ & $0.223^{+0.007}_{-0.007}$ \\
1998 DK$_{36}$ & $1.00^{+0.00}_{-0.43}$ & $0.00^{+0.43}_{-0.00}$ & $0.00^{+0.43}_{-0.00}$ & $\sim0.050$ \\
2010 XB$_{11}$ & $0.13^{+0.01}_{-0.01}$ & $0.52^{+0.02}_{-0.02}$ & $0.34^{+0.02}_{-0.02}$ & $0.213^{+0.007}_{-0.007}$ \\
2012 VE$_{46}$ & $0.20^{+0.01}_{-0.01}$ & $0.64^{+0.01}_{-0.01}$ & $0.16^{+0.01}_{-0.01}$ & $0.186^{+0.004}_{-0.004}$ \\
2013 JX$_{28}$ & $0.14^{+0.01}_{-0.01}$ & $0.63^{+0.01}_{-0.01}$ & $0.23^{+0.01}_{-0.01}$ & $0.202^{+0.004}_{-0.004}$ \\
2013 TQ$_{5}$ & $0.13^{+0.01}_{-0.01}$ & $0.59^{+0.01}_{-0.01}$ & $0.28^{+0.01}_{-0.01}$ & $0.209^{+0.004}_{-0.004}$ \\
2014 FO$_{47}$ & $0.18^{+0.01}_{-0.01}$ & $0.67^{+0.01}_{-0.01}$ & $0.15^{+0.01}_{-0.01}$ & $0.188^{+0.004}_{-0.004}$ \\
2015 DR$_{215}$ & $0.27^{+0.02}_{-0.02}$ & $0.69^{+0.02}_{-0.02}$ & $0.04^{+0.01}_{-0.01}$ & $0.164^{+0.005}_{-0.005}$ \\
2017 XA$_{1}$ & $0.68^{+0.01}_{-0.01}$ & $0.32^{+0.01}_{-0.01}$ & $0.00^{+0.00}_{-0.00}$ & $0.098^{+0.002}_{-0.002}$ \\
2018 JB$_{3}$ & $0.11^{+0.03}_{-0.02}$ & $0.38^{+0.04}_{-0.04}$ & $0.51^{+0.04}_{-0.04}$ & $0.235^{+0.015}_{-0.014}$ \\
2019 AQ$_{3}$ & $0.12^{+0.04}_{-0.04}$ & $0.53^{+0.06}_{-0.06}$ & $0.35^{+0.06}_{-0.05}$ & $0.217^{+0.022}_{-0.019}$ \\
2019 LF$_{6}$ & $0.14^{+0.03}_{-0.03}$ & $0.49^{+0.04}_{-0.04}$ & $0.37^{+0.04}_{-0.04}$ & $0.216^{+0.015}_{-0.015}$ \\
2020 HA$_{10}$ & $0.12^{+0.02}_{-0.02}$ & $0.44^{+0.04}_{-0.04}$ & $0.44^{+0.04}_{-0.04}$ & $0.226^{+0.014}_{-0.014}$ \\
2020 OV$_{1}$ & $0.12^{+0.01}_{-0.01}$ & $0.50^{+0.02}_{-0.02}$ & $0.38^{+0.02}_{-0.02}$ & $0.220^{+0.007}_{-0.007}$ \\
2021 BS$_{1}$ & $0.12^{+0.01}_{-0.01}$ & $0.50^{+0.02}_{-0.02}$ & $0.38^{+0.02}_{-0.02}$ & $0.220^{+0.007}_{-0.007}$ \\
2021 LJ$_{4}$ & $0.18^{+0.01}_{-0.01}$ & $0.62^{+0.01}_{-0.01}$ & $0.20^{+0.01}_{-0.01}$ & $0.193^{+0.004}_{-0.004}$ \\
2021 PB$_{2}$ & $0.11^{+0.01}_{-0.01}$ & $0.51^{+0.02}_{-0.02}$ & $0.38^{+0.02}_{-0.02}$ & $0.222^{+0.007}_{-0.007}$ \\
2021 PH$_{27}$ & $0.10^{+0.05}_{-0.04}$ & $0.48^{+0.07}_{-0.07}$ & $0.42^{+0.07}_{-0.07}$ & $0.227^{+0.025}_{-0.025}$ \\
2021 VR$_{3}$ & $0.08^{+0.02}_{-0.02}$ & $0.57^{+0.03}_{-0.03}$ & $0.35^{+0.03}_{-0.03}$ & $0.223^{+0.011}_{-0.011}$ \\
2022 BJ$_{8}$ & $0.13^{+0.01}_{-0.01}$ & $0.54^{+0.01}_{-0.01}$ & $0.34^{+0.01}_{-0.01}$ & $0.217^{+0.004}_{-0.004}$ \\
2023 EL & $0.13^{+0.01}_{-0.01}$ & $0.51^{+0.01}_{-0.01}$ & $0.37^{+0.01}_{-0.01}$ & $0.220^{+0.004}_{-0.004}$ \\
2023 EY$_{2}$ & $0.14^{+0.01}_{-0.01}$ & $0.56^{+0.02}_{-0.02}$ & $0.30^{+0.02}_{-0.02}$ & $0.209^{+0.007}_{-0.007}$ \\
2023 KQ$_{5}$ & $0.40^{+0.19}_{-0.17}$ & $0.60^{+0.17}_{-0.19}$ & $0.00^{+0.17}_{-0.00}$ & $0.140^{+0.062}_{-0.039}$ \\
2023 WK$_{3}$ & $0.23^{+0.01}_{-0.01}$ & $0.71^{+0.01}_{-0.01}$ & $0.06^{+0.01}_{-0.00}$ & $0.171^{+0.004}_{-0.002}$ \\
2024 UM$_{9}$ & $0.29^{+0.01}_{-0.01}$ & $0.70^{+0.01}_{-0.01}$ & $0.01^{+0.00}_{-0.00}$ & $0.157^{+0.002}_{-0.002}$ \\
2024 WD$_{19}$ & $0.68^{+0.03}_{-0.03}$ & $0.32^{+0.03}_{-0.03}$ & $0.00^{+0.00}_{-0.00}$ & $0.098^{+0.006}_{-0.006}$ \\
2025 GN$_{1}$ & $0.07^{+0.06}_{-0.04}$ & $0.57^{+0.09}_{-0.09}$ & $0.37^{+0.09}_{-0.08}$ & $0.228^{+0.033}_{-0.030}$ \\
2025 LP & $0.11^{+0.01}_{-0.01}$ & $0.55^{+0.02}_{-0.02}$ & $0.34^{+0.01}_{-0.01}$ & $0.218^{+0.005}_{-0.005}$ \\
2025 SC$_{79}$ & $0.10^{+0.01}_{-0.01}$ & $0.53^{+0.02}_{-0.02}$ & $0.37^{+0.02}_{-0.02}$ & $0.222^{+0.007}_{-0.007}$ \\
2026 BX$_{4}$ & $0.44^{+0.01}_{-0.01}$ & $0.56^{+0.01}_{-0.01}$ & $0.00^{+0.00}_{-0.00}$ & $0.134^{+0.002}_{-0.002}$ \\
\textbf{(3200) Phaethon} & $0.07^{+0.06}_{-0.04}$ & $0.54^{+0.09}_{-0.09}$ & $0.39^{+0.09}_{-0.09}$ & $0.229^{+0.033}_{-0.033}$/$0.16^{+0.02}_{-0.02}$$^{***}$\\
\textbf{(155140) 2005 UD} & $0.14^{+0.02}_{-0.02}$ & $0.50^{+0.03}_{-0.03}$ & $0.36^{+0.03}_{-0.03}$ & $0.215^{+0.011}_{-0.011}$/$0.14^{+0.09}_{-0.09}$$^{***}$ \\
\textbf{(225416) 1999 YC} & $0.15^{+0.02}_{-0.02}$ & $0.48^{+0.03}_{-0.03}$ & $0.37^{+0.03}_{-0.03}$ & $0.215^{+0.011}_{-0.011}$/$0.09^{+0.03}_{-0.03}$$^{****}$  \\
\hline
\end{tabular}
\vspace{0.5em}

\begin{minipage}{0.95\linewidth}
\footnotesize
\textbf{Notes.} Values in the three albedo-category columns are the average
probabilities in each category, with upper and lower 1-$\sigma$ uncertainties.
The final column gives the weighted model albedo, $p_V$, with its upper and
lower 1-$\sigma$ uncertainties. (1) Low albedo: $p_V<0.1$ probability; (2) medium albedo:
$0.1\leq p_V<0.3$ probability; (3) high albedo: $p_V\geq0.3$ probability; and (4) weighted model albedo: $\tilde p_V=0.05 \, P_{\rm low}+0.20 \, P_{\rm intermediate}+0.30 \, P_{\rm high}$, *albedo estimate from \citet[][]{Deleon2024}, **albedo estimate from \citet[][]{Bolin2022IVO}, ***albedos from \citet[][]{Masiero2019}, ****albedos from \citet[][]{Kasuga2022}.
\end{minipage}
\end{table}

\end{document}